\documentclass[pra,amsmath,amsfonts,onecolumn,superscriptaddress,showpacs, showkeys,preprint]{revtex4-1}

\usepackage{epsf,epsfig}
\usepackage[psamsfonts]{amssymb}
\usepackage{amsmath}
\usepackage{graphicx}
\usepackage{color,soul}
\usepackage{xcolor}

 \newcommand{\ket}[1]{|#1\rangle}
 \newcommand{\bra}[1]{\langle #1|}

  \newcommand{\inp}{{\mathrm in}}
   \newcommand{\out}{{\mathrm out}}

\newcommand{\Tr}{{\mathrm {Tr}}}

\begin{document}
\title{  
Two-qubit universal and state-dependent quantum cloning machine and  quantum correlation broadcasting  \\
}
\author{Azam Kheirollahi\footnote{a.kheirollahi@sci.ui.ac.ir}}\affiliation{Department of Physics, University of Isfahan,
 Isfahan, Iran}
 \author{Hamidreza Mohammadi\footnote{hr.mohammadi@sci.ui.ac.ir}}\affiliation{Department of Physics, University of Isfahan,
 Isfahan, Iran}
\affiliation{Quantum Optics Group, University of Isfahan,
 Isfahan, Iran}
\author{Seyed Javad Akhtarshenas\footnote{akhtarshenas@um.ac.ir}}
\affiliation{Department of Physics, Ferdowsi University of Mashhad,
 Mashhad, Iran}

%\author{}
%\affiliation{Department of Physics, Ferdowsi University of Mashhad,
% Mashhad, Iran}
%\affiliation{Quantum Optics Group, University of Isfahan,
% Isfahan, Iran}

\begin{abstract}
Due to the axioms of quantum mechanics, perfect cloning of an unknown quantum state is impossible. But since imperfect cloning is still possible, a question arises: "Is there an optimal quantum cloning machine?" Buzek and Hillery answer to this question and construct their famous B-H quantum cloning machine. The B-H machine clones state of an arbitrary single qubit in optimal manner and hence it is universal. Generalizing this machine for two-qubit system is straightforward, but this procedure does not preserve quantum correlation existing in bipartite state in optimal manner and also, during this procedure, this machine loses its universality and becomes a state-dependent cloning machine. In this paper we propose an optimal universal local quantum state cloner for two qubit systems. Also we present two classes of state-dependent local quantum copying machine. Furthermore, we investigate local broadcasting of two aspects of quantum  correlations, i.e., quantum entanglement and quantum discord defined, respectively, within the entanglement-separability paradigm and from an information-theoretic perspective. The results show that although quantum correlation is in general  very fragile during broadcasting procedure, quantum discord is broadcasted more robustly than quantum entanglement.
\end{abstract}

\maketitle

\section{Introduction}
In quantum world there is no way to duplicate, perfectly, a system that is initially in an unknown quantum state. This is called \textit{no-cloning theorem} \cite{Dieks1982, WoottersZurek1982}. The no-cloning theorem states that there does not exist an ideal quantum copying machine which can take two distinct non-orthogonal quantum states ($ \ket{\psi},\ket{\phi} $) into states ($ \ket{\psi}\otimes\ket{\psi},\ket{\phi}\otimes\ket{\phi} $), respectively. Consider two orthogonal states $ \ket{0} $ and $ \ket{1} $, we can construct a perfect cloning procedure as a unitary transformation, possibly involving a machine (an ancilla) \cite{kraus1983} such that
\begin{eqnarray}\label{cloning}
U\ket{0}\ket{\textbf{b}}\ket{M} &\rightarrow & \ket{0}\ket{0}\ket{M_{0}},\\ \nonumber
U\ket{1}\ket{\textbf{b}}\ket{M} &\rightarrow & \ket{1}\ket{1}\ket{M_{1}},
\end{eqnarray}
where $ \ket{\textbf{b}} $ is the blank state,  $ \ket{M} $ is the initial state of the ancilla, and $ \ket{M_{i}}$ ($i=0,1$) denote the output ancilla states.
This implies that
\begin{equation}
U(\ket{0}+\ket{1})\ket{\textbf{b}}\ket{M}\rightarrow \ket{0}\ket{0}\ket{M_{0}}+\ket{1}\ket{1}\ket{M_{1}},
\end{equation}
which is not equal to $ (\ket{0}+\ket{1})(\ket{0}+\ket{1})\ket{M_{0+1}} $. This non-equality arises from the linearity of quantum mechanics.
%\textcolor{blue}{In 1996 Barnum et.al. mentioned for noncommuting mixed states, the impossibility of the perfect cloning \cite{Barnum1996}.}
In spite of the no-cloning theorem, we can construct a quantum cloning machine (QCM) that copy an unknown quantum state approximately. Buzek et.al.  \cite{Buzek1996} introduced a universal quantum cloning machine (UQCM) which its copying quality remains same for all input states. The fidelity of this machine is equal $ 5/6 $, regardless of its input state. The fidelity $F$ of a QCM is defined as "how its output state $\rho_{\out}$ is similar to input state $\rho_{\inp}$ of machine?". It was also shown that the B-H UQCM machine is optimal \cite{Gisin1997, Bruss1998, Gisin1999}. Buzek and Hillery  extended their optimal UQCM to clones quantum states in arbitrary dimensional Hilbert space \cite{Buzek1998}.
On the other hand, there are quantum cloning machines which their fidelity  depends on the parameters of the input state, i.e., the so-called "state-dependent QCM." The Wootters-Zurek (W-Z) quantum cloning machine is the first example that its copying quality depends on the input state \cite{WoottersZurek1982}. Various state-dependent QCMs were also proposed \cite{Bruss2000, Niu1999, Durt2004, Cerf2002}.

The above mentioned QCMs act only on a single party systems. But broadcasting of quantum correlation is important and essential for quantum information and computation processing \cite{Buzek1997, Adhikari2006, Bandyopadhyay1999}. Broadcasting the quantum correlation needs the construction of QCMs for multi-partite systems. A simple way is to employ B-H machine for coping state of each party, locally. Calculations show that this machine becomes a state-dependent QCM and also it does not preserve quantum correlations existing in the bipartite system, in an optimal manner. Quantum information processing uses quantum correlation as a resource to communicate between to distant parties, hence local broadcasting is more important and useful. Thus non-local B-H machine is not proper for operational quantum information processing.

In this paper we propose a local UQCM for two qubit pure states. We also introduce two classes of state-dependent QCMs, by identifying our machine with Pauli channel operations. In this manner, we construct our cloners by comparing the fidelity between input and output states of a symmetric unitary transformation with the fidelity between input and output states of the Pauli channel operations. A comparison of the ability of QCMs in  broadcasting different type of quantum correlations is also investigated. Quantum entanglement \cite{Einstein1935,Schrodinger1935} and quantum discord \cite{Ollivier2001, Henderson2001} are two aspects of quantum correlations defined, respectively, within the entanglement-separability paradigm and from an information-theoretic perspective. More precisely, for the two-qubit pure and Werner states we  study the robustness of these quantum correlations under broadcasting.  We use entanglement of formation \cite{Wootters1998} and original quantum discord \cite{Ollivier2001, Henderson2001} for measuring the amount of quantum correlations existing in output states of cloners. Our results reveals that quantum correlation is in general  very fragile during broadcasting procedure, and quantum discord is broadcasted  more robustly than quantum entanglement.

The organization of this paper is as follow. In Sec. II, we introduce local and non-local B-H quantum copying machines and then we construct our local universal quantum copying machine for two-qubit pure states in subsection A. We then present two classes of local state-dependent quantum cloning machines for copying two-qubit states. In Section III, we study  broadcasting of quantum correlations via these QCMs. The paper is concluded in section IV with a brief discussion.

\section{Quantum copying machines}
Generalized UQCM introduced by Buzek and Hillery which broadcasts quantum states in $M$-dimensional Hilbert space can be modeled by the following unitary transformation \cite{Buzek1998}
\begin{equation}\label{BH nonlocal transformation}
U_{BH}\ket{\psi_{i}}\ket{\textbf{b}}\ket{X}\rightarrow c \ket{\psi_{i}}\ket{\psi_{i}}\ket{X_{i}}+d \sum_{j\neq i}^{M}\left(\ket{\psi_{i}}\ket{\psi_{j}}+\ket{\psi_{j}}\ket{\psi_{i}}\right)\ket{X_{j}},
\end{equation}
where$\ket{\textbf{b}}$  refers to blank state, $ \ket{X} $ denotes the initial state of the ancilla (machine), $ \ket{X_{i}} $ refer to output ancilla states and the cofficients $ c=\sqrt{\frac{2}{M+1}} $ and $ d=\sqrt{\frac{1}{2(M+1)}} $ are real. For the case of  qubits, the dimension of Hilbert space is $ M=2 $. So, the basis vectors of this Hilbert space are $ \ket{\psi_{1}}=\ket{0} $ and $ \ket{\psi_{2}}=\ket{1} $. Also, the coefficients $ c $ and $ d $ are  $ c=\sqrt{\frac{2}{3}} $ and $ d=\sqrt{\frac{1}{6}} $. Therefore, the Buzek-Hillery UQCM for a given input qubit state is \cite{Buzek1996}
\begin{eqnarray}\label{BH transformation}
U_{BH}\ket{0}\ket{0}\ket{X}&=&\sqrt{\frac{2}{3}} \ket{0}\ket{0}\ket{A}+\sqrt{\frac{1}{6}} \left(\ket{0}\ket{1}+\ket{1}\ket{0}\right)\ket{A_{\perp}}, \\ \nonumber
U_{BH}\ket{1}\ket{0}\ket{X}&=&\sqrt{\frac{2}{3}} \ket{1}\ket{1}\ket{A_{\perp}}+\sqrt{\frac{1}{6}} \left(\ket{1}\ket{0}+\ket{0}\ket{1}\right)\ket{A}.
\end{eqnarray}
The fidelity B-H transformation defined in Eq.\eqref{BH transformation} for copying of a qubit state $ \ket{\psi}=\alpha\ket{0}+\beta\ket{1} $ is equal to $ 5/6 $ for all $\alpha$ and $\beta$.

Now, let us assume that two distant parties $ A $ and $ B $ share  two-qubit pure state
 \begin{equation}\label{pure two qubit state}
\ket{\psi}=\alpha\ket{00}+\beta\ket{11},
\end{equation}
with $ \alpha, \beta \in \mathbb{R} $ and $ \alpha^{2}+\beta^{2}=1 $. Suppose, the first qubit belongs to $ A $ and the second qubit belongs to $ B $. We consider the case that each of these two parties $ A $ and $ B $ perform B-H transformation defined in Eq. \eqref{BH transformation}, locally. By tracing over the machine vectors and after tracing over one of the two parties we get
\begin{eqnarray}\label{local}
\rho_{\out}^{\textrm{BH-l}}&=&\frac{1}{36}\left[\left(25\alpha ^{2}+\beta ^{2}\right)\ket{00}\bra{00}+\left(\alpha ^{2}+25\beta^{2}\right)
\ket{11}\bra{11}\right.\\\nonumber
 &+& \left. 5\left(\ket{01}\bra{01}+\ket{10}\bra{10}\right)
+16\alpha\beta \left(\ket{00}\bra{11}+\ket{11}\bra{00}\right)\right].
\end{eqnarray}
The fidelity between the input  and  output states is
  \begin{equation}\label{ fidelity2}
F(\alpha^{2})=\bra{\psi}\rho_{\out}^{\textrm{BH-l}}\ket{\psi}=\frac{1}{36}\left(16\alpha^{4}-16\alpha^{2}+25\right),
\end{equation}
which, clearly, reveals that the local B-H machine when acting on a two-qubit state becomes state-dependent and hence loses its universality. The average fidelity of this machine is given by
\begin{equation}\label{ average fidelity2}
\overline{F}=\int_{0}^{1}{F(\alpha^{2})d\alpha^{ 2}}=0.62.
\end{equation}
To overcome this problem, Hillery and Buzek propose their non-local UQCM. They considered a two-qubit state as a four-dimensional Hilbert space, i.e., substituting $M=4$ in Eq. \eqref{BH nonlocal transformation}. The basis vectors of this Hilbert space are $ \ket{\psi_{1}}=\ket{00} $, $ \ket{\psi_{2}}=\ket{01} $, $ \ket{\psi_{3}}=\ket{10} $ and $ \ket{\psi_{4}}=\ket{11} $. In this case, the coefficients $ c $ and $ d $ take the values $ c=\sqrt{\frac{2}{5}} $ and $ d=\sqrt{\frac{1}{10}} $. The unitary transformation of such machine is not local with respect to the A and B parties. This fact reduces the operationality of this UQCM. By tracing over the machine vectors and after tracing over one of the two parties, we can write
 \begin{equation}\label{nonl}
\rho_{\out}^{\textrm{BH-nl}}=\frac{1}{10}\left[\mathbb{I}\otimes \mathbb{I}+6\left(\alpha ^{2}\ket{00}\bra{00}+\beta ^{2}\ket{11}\bra{11}\right)
+6\alpha\beta \left(\ket{00}\bra{11}+\ket{11}\bra{00}\right)\right].
\end{equation}
where $\mathbb{I}$ is the $2$-dimensional unit  matrix of each subsystem. In this case the fidelity between input state $ \ket{\psi}=\alpha\ket{00}+\beta\ket{11} $ and output state is given by
\begin{equation}\label{fidelitynonl}
F=\langle\psi\vert\rho_{\out}^{\textrm{BH-nl}}\vert\psi\rangle =0.7.
\end{equation}
As we see, the fidelity of non-local case does not depend on the parameters of input state and hence the map is universal.
In the following, we propose a universal local broadcasting machine for two-qubit pure states and then we explore about state-dependent QCMs for two-qubit systems.

\subsection{ Universal quantum cloning machine for two-qubit states}
A UQCM can be achieved by the following general unitary operation
 \begin{eqnarray}\label{G unitary transformation}
U\ket{0}\ket{0}\ket{X}&=&a\ket{0}\ket{0}\ket{A}+b_{1}\ket{0}\ket{1}\ket{B_{1}}+b_{2}\ket{1}\ket{0}\ket{B_{2}}
+c\ket{1}\ket{1}\ket{C},\\\nonumber
U\ket{1}\ket{0}\ket{X}&=&a^{\prime}\ket{1}\ket{1}\ket{A^{\prime}}+b^{\prime}_{1}\ket{1}\ket{0}
\ket{B^{\prime}_{1}}+b^{\prime}_{2}\ket{0}\ket{1}\ket{B^{\prime}_{2}}
+c^{\prime}\ket{0}\ket{0}\ket{C^{\prime}},
\end{eqnarray}
where $ \ket{X} $ denotes the initial state of the ancilla  and $ \ket{A},\ket{B_{i}},\ket{C},... $ refer to the output ancilla states. Due to unitarity, the only condition that we have on these states is that they are orthonormal, so that  the coefficients $ a,b_{i},c,... $ satisfy the normalization condition:
 \begin{eqnarray}\label{G normalization condition}
\vert a\vert^{2}+\vert b_{1}\vert^{2}+\vert b_{2}\vert^{2}+\vert c\vert^{2}=\vert a^{\prime}\vert^{2}+\vert b^{\prime}_{1}\vert^{2}+\vert b^{\prime}_{2}\vert^{2}+\vert c^{\prime}\vert^{2}=1,
\end{eqnarray}
and the orthogonalization condition:
 \begin{equation}\label{G orthogonality condition}
a^{\ast}c^{\prime}\langle A\vert C^{\prime}\rangle +b_{2}^{\ast}b^{\prime}_{1}\langle{B_{2}}\vert{B^{\prime}_{1}}\rangle +
b_{1}^{\ast}b^{\prime}_{2}\langle{B_{1}}\vert{B^{\prime}_{2}}\rangle +c^{\ast}a^{\prime}\langle{C}\vert{A^{\prime}}\rangle =0.
\end{equation}\\
The symmetry of transformation implies that $ \rho_{a_{1}b_{1}}= \rho_{a_{2}b_{2}}$, where  $\rho_{a_{i}b_{i}}=\Tr_{\bar{a}_i,\bar{b}_i}(\rho_{\out})$ and the  trace is taken over all subsystems except $a_i$ and $b_i$. Hence, by applying this transformation on a general pure state $ \ket{\psi}=\alpha\ket{00}+\beta\ket{11} $ and after tracing over ancillary states and states of subsystem $ a_{2}b_{2} $ or $ a_{1}b_{1} $, we obtain $ b_{1}=b_{2}=b $, $ b^{\prime}_{1}=b^{\prime}_{2}=b^{\prime} $, $ \ket{B_{1}}= \ket{B_{2}}= \ket{B} $ and $ \ket{B^{\prime}_{1}}=\ket{B^{\prime}_{2}}=\ket{B^{\prime}} $.
Therefore, we can write
 \begin{eqnarray}\label{sym unitary transformation}
U\ket{0}\ket{0}\ket{X}&=&\vert a\vert e^{i\delta_{a}}\ket{0}\ket{0}\ket{A}+\vert b\vert e^{i\delta_{b}}\left.(\ket{0}\ket{1}+\ket{1}\ket{0}\right)\ket{B}
+\vert c\vert e^{i\delta_{c}}\ket{1}\ket{1}\ket{C}, \\\nonumber
U\ket{1}\ket{0}\ket{X}&=&\vert a^{\prime}\vert e^{i\delta_{a^{\prime}}}\ket{1}\ket{1}\ket{A^{\prime}}+\vert b^{\prime}\vert e^{i\delta_{b^{\prime}}}\left(\ket{1}\ket{0}+\ket{0}\ket{1}\right)\ket{B^{\prime}}
+\vert c^{\prime}\vert e^{i\delta_{c^{\prime}}}\ket{0}\ket{0}\ket{C^{\prime}}.
\end{eqnarray}
Without loss of generality we can absorb phases $ \lbrace\delta_{a},\delta_{b}, ....\rbrace $ to states $ \lbrace\ket{A}, \ket{B},...\rbrace $, so we obtain
 \begin{eqnarray}\label{sym unitary transformation1}
U\ket{0}\ket{0}\ket{X}&=&\vert a\vert \ket{0}\ket{0}\ket{A}+\vert b\vert \left(\ket{0}\ket{1}+\ket{1}\ket{0}\right)\ket{B}
+\vert c\vert \ket{1}\ket{1}\ket{C},\\\nonumber
U\ket{1}\ket{0}\ket{X}&=&\vert a^{\prime}\vert \ket{1}\ket{1}\ket{A^{\prime}}+\vert b^{\prime}\vert \left(\ket{1}\ket{0}+\ket{0}\ket{1}\right)\ket{B^{\prime}}
+\vert c^{\prime}\vert \ket{0}\ket{0}\ket{C^{\prime}}.
\end{eqnarray}
Also, because of symmetry of unitary transformation under exchanges between $ \ket{0}\leftrightarrow\ket{1} $ we must have
 \begin{equation}\label{exchange01}
\vert a\vert =\vert a^{\prime}\vert,\quad\quad\vert b\vert =\vert b^{\prime}\vert,\quad\quad\vert c\vert =\vert c^{\prime}\vert.
\end{equation}
Now consider a universal $1 \rightarrow 2$ quantum state cloner that takes a completely unknown pure two-qubit state $ \ket{\psi}=\alpha\ket{00}+\beta\ket{11} $, as input state, and generates two copy of it as the output. The optimality and universality implies that the fidelity $F$ between  $\rho_{\inp}=\ket{\psi}\bra{\psi} $ and $\rho_{a_{1}b_{1}}$ has maximum value and also is independent of the initial parameters $\alpha$ and $\beta$. The procedure of maximization of $F$ subject to the constrains defined in Eqs. \eqref{G normalization condition} and \eqref{G orthogonality condition} is done by using Lagrange multiplier method and is given in Appendix A. Accordingly, the optimal symmetric and universal quantum cloning machine for local cloning of two-qubit pure states could be achieve for $|a|=\frac{1}{\sqrt{2}}$,  $ |b|=\frac{1}{2} $, $ |c|=0 $ hence
 \begin{eqnarray}\label{opt unitary transformation}
U\ket{0}\ket{0}\ket{X}&=&\frac{1}{\sqrt{2}} \ket{0}\ket{0}\ket{A}+\frac{1}{2} \left(\ket{0}\ket{1}+\ket{1}\ket{0}\right)\ket{A_{\perp}}, \\ \nonumber
U\ket{1}\ket{0}\ket{X}&=&\frac{1}{\sqrt{2}} \ket{1}\ket{1}\ket{A_{\perp}}+\frac{1}{2} \left(\ket{1}\ket{0}+\ket{0}\ket{1}\right)\ket{A}.
\end{eqnarray}
Performing the above state independent QCM  on general pure state $ \ket{\psi}=\alpha\ket{00}+\beta\ket{11} $ gives us
\begin{eqnarray}\label{pure independent}
\rho_{a_1,b_1}&=&\rho_{a_2,b_2}=\frac{1}{16}\left[
\left(9\alpha^{2}+\beta^{2}\right)\ket{00}\bra{00}+\left(\alpha^{2}+9\beta^{2}\right)\ket{11}\bra{11}\right.\\\nonumber
&+& \left. 3\left(\ket{01}\bra{01}+\ket{10}\bra{10}\right)+8\alpha\beta\left(\ket{00}\bra{11}+\ket{11}\bra{00}\right)\right].
\end{eqnarray}
So, the fidelity of this universal quantum copying machine is $ F=(\vert a\vert^{2}+\vert b\vert^{2})^{2}=\frac{9}{16} $, regardless of values of $\alpha$ and $\beta$.

\subsection{State-dependent cloning machine}
Can we construct a cloning machine for an special class of bipatite quantum states, with higher degree of fidelity? The answer is positive and leads to construction of  the state-dependent QCMs for bipartite system. In this subsection we look for state-dependent cloning machine for two-qubit quantum states. In this way we can construct our QCM by comparing the fidelity between input and output states of the map \eqref{sym unitary transformation1} with the fidelity between input and output states of the Pauli channel operations.  The operation on an arbitrary two-qubit input state $ \rho_{\inp} $ performed by the Pauli channel is given by \cite{Macchiavello2003, Petz2008}
\begin{equation}\label{General Pauli channel}
\rho =s\rho_{\inp} +\sum _{i=1}^{3}p_{i}\left(\sigma_{i}\otimes \mathbb{I}\right)\rho_{\inp}\left(\sigma_{i}\otimes \mathbb{I}\right)+\sum _{i=1}^{3}q_{i}\left(\mathbb{I}\otimes\sigma_{i}\right)\rho_{\inp}\left(\mathbb{I}\otimes\sigma_{i}\right)+\sum _{i,j=1}^{3}t_{i,j}\left(\sigma_{i}\otimes\sigma_{j}\right)\rho_{\inp}\left(\sigma_{i}\otimes\sigma_{j}\right),
\end{equation}
where $\mathbb{I}$ is the $2$-dimensional unit  matrix of each subsystem, $ \sigma_{i}=\lbrace \sigma_{1}, \sigma_{2}, \sigma_{3}\rbrace $ are the Pauli matrices and $ s+\sum _{i=1}^{3}p_{i}+\sum _{i=1}^{3}q_{i}+\sum _{i,j=1}^{3}t_{i,j}=1 $. In the following we consider two special cases

{\it One-Pauli channel.---}
First  we use the "one-Pauli" channel, i.e., $ p_{3}=q_{3}=t_{33}=(1-s)/3 $, we have
\begin{equation}\label{1Pauli channel}
\rho =s\rho_{\inp} +\left(\frac{1-s}{3}\right)\left\lbrace\left(\sigma_{3}\otimes \mathbb{I}\right)\rho_{\inp}\left(\sigma_{3}\otimes \mathbb{I}\right)+\left(\mathbb{I}\otimes\sigma_{3}\right)\rho_{\inp}\left(\mathbb{I}\otimes\sigma_{3}\right)
+\left(\sigma_{3}\otimes\sigma_{3}\right)\rho_{\inp}\left(\sigma_{3}\otimes\sigma_{3}\right)\right\rbrace.
\end{equation}
For this channel, if we use a two-qubit pure input state $ \rho_{\inp}=\ket{\psi}\bra{\psi} $ with $ \ket{\psi}=\alpha\ket{00}+\beta\ket{11} $, we get
\begin{equation}\label{1Pauli channel pure}
\rho =\alpha^{2}\ket{00}\bra{00}+\frac{4s-1}{3}\alpha\beta\left(\ket{00}\bra{11}+\ket{11}\bra{00}\right)+\beta^{2}\ket{11}\bra{11}.
\end{equation}
In this case fidelity is $ F_{ch}=\langle \psi\vert \rho\vert\psi\rangle=1+\frac{8}{3}(s-1)\alpha^{2}\beta^{2} $ which, clearly, depends  on the input state parameters and takes its maximum value  1 for product input states, \textit{i.e.}, when $ \alpha=0 $ or $ \alpha=1 $.
In the following, we apply the symmetric quantum cloning machine, defined by the map presented in Eq. \eqref{sym unitary transformation1}, on $ \rho_{\inp}=\ket{\psi}\bra{\psi} $. Due to symmetric property of machine, after tracing over ancillary states and states of subsystems $ a_{1}b_{1} $ or $ a_{2}b_{2} $, we find $ \rho_{a_{1}b_{1}}=\rho_{a_{2}b_{2}}=\rho_{out} $. After that we calculate the fidelity between input state and output state of this machine \textit{i.e.}, $ F=\langle \psi\vert \rho_{out}\vert\psi\rangle $. \\
For the case $\alpha=0$, a comparison between two fidelities $F$ and $F_{ch}$ gives $ \vert a\vert ^{2}+\vert b\vert ^{2}=1 $. Also the unitarity condition implies that  $ \vert a\vert ^{2}+2\vert b\vert ^{2}+\vert c\vert ^{2}=1 $. So we have $ \vert b\vert ^{2}+\vert c\vert ^{2}=0 $. Due to reality and positivity of $ \vert b\vert $ and $ \vert c\vert $, we conclude that $ \vert b\vert =\vert c\vert =0 $ and $ \vert a\vert =1 $. Finally, the fidelity is  $ F=\langle \psi\vert \rho_{out}\vert\psi\rangle =\alpha ^{4}+\beta ^{4}=1-2\alpha^{2}\beta^{2} $, and the average fidelity is
\begin{equation}\label{1Pauli average fidelity}
\overline{F}=\int_{0}^{1}{F(\alpha^{2})d\alpha^{ 2}}=0.66.
\end{equation}

{\it Two-Pauli channel.---}
As the second case, if we set  $p_{i}=q_{i}=t_{ij}=0$  for $i,j=2 $,  we get the  two-Pauli channel, that is
\begin{equation}\label{2Pauli channel}
\rho =s\rho_{\inp} +(\frac{1-s}{8})\lbrace\sum_{i=1,3}(\sigma_{i}\otimes \mathbb{I})\rho_{\inp}(\sigma_{i}\otimes \mathbb{I})+\sum_{i=1,3}(\mathbb{I}\otimes\sigma_{i})\rho_{\inp}(\mathbb{I}\otimes\sigma_{i})+\sum_{i,j=1,3}(\sigma_{i}\otimes\sigma_{j})\rho_{\inp}(\sigma_{i}\otimes\sigma_{j})\rbrace.
\end{equation}
The output of this channel for the two-qubit pure state $ \ket{\psi}=\alpha\ket{00}+\beta\ket{11} $, as the input state $ \rho_{\inp} $, is
\begin{eqnarray}\label{2Pauli channel pure}
\rho&=&\frac{1}{8}(1+2\alpha^{2}+(6\alpha ^{2}-1)s)\ket{00}\bra{00}+s\alpha\beta(\ket{00}\bra{11}+\ket{11}\bra{00})\\\nonumber
&+&\frac{1-s}{4}(\ket{01}\bra{01}+\ket{10}\bra{10})+\frac{1}{8}(1+2\beta ^{2}+(6\beta ^{2}-1)s)\ket{11}\bra{11}.
\end{eqnarray}
For this case the fidelity between input state and output state of the channel is $ F_{ch}=\langle \psi\vert \rho\vert\psi\rangle=\frac{1}{8}(3+5s-4(1-s)\alpha^{2}\beta^{2}) $. As we see the fidelity depends on the input state parameters.\\
 Now, we consider the fidelity between input state and output state of the map presented in Eq. \eqref{sym unitary transformation1}\textit{i.e.}, $ F=\langle \psi\vert \rho_{out}\vert\psi\rangle $. Our aim is to construct this QCM by comparing these two fidelities. For the case of $ \alpha =0 $, comparison gives us $ \frac{3+5s}{8}=(\vert a\vert ^{2}+\vert b\vert ^{2})^{2} $ and for another values of $ \alpha $ leads us to $ \frac{1}{4}(1-s)=2(\vert a\vert ^{2}+\vert b\vert ^{2})-1-\vert a\vert ^{2}\vert b\vert ^{2}Re(\langle B^{\prime}\vert A\rangle +\langle A^{\prime}\vert B\rangle )^{2} $. The procedure of maximizing the fidelity with respect to the parameter $s$ is given in Appendix B.
The results of Appendix B introduce a class of state-dependent quantum cloning machine (we call it two-Pauli-like cloning machine) for local cloning of two-qubit states as
 \begin{eqnarray}\label{two-Pauli-like transformation1}
U\ket{0}\ket{0}\ket{X}&=&\sqrt{\frac{4+\sqrt{79}}{21}} \ket{0}\ket{0}\ket{A}+\sqrt{\frac{17-\sqrt{79}}{42}} (\ket{0}\ket{1}+\ket{1}\ket{0})\ket{A_{\perp}},\\\nonumber
U\ket{1}\ket{0}\ket{X}&=&\sqrt{\frac{4+\sqrt{79}}{21}} \ket{1}\ket{1}\ket{A_{\perp}}+\sqrt{\frac{17-\sqrt{79}}{42}} (\ket{1}\ket{0}+\ket{0}\ket{1})\ket{A}.
\end{eqnarray}
By performing the two-Pauli-like cloning machine \eqref{two-Pauli-like transformation1} on a general pure state $ \ket{\psi}=\alpha\ket{00}+\beta\ket{11} $ we have
\begin{eqnarray}\label{pure 2pauli} \rho_{\out}&=&\left(0.6510\alpha^{2}+0.0337\beta^{2}\right)\ket{00}\bra{00}+\left(0.0337\alpha^{2}+0.6510\beta^{2}\right)\ket{11}\bra{11}\\\nonumber
&+&0.1558\left(\ket{01}\bra{01}+\ket{10}\bra{10}\right)
+0.4741\alpha\beta\left(\ket{00}\bra{11}+\ket{11}\bra{00}\right).
\end{eqnarray}
Therefore, for this cloning machine, the fidelity is $ F=\frac{1}{8}(5.205-2.236\alpha^{2}\beta^{2}) $, and the average fidelity is
\begin{equation}\label{2Pauli average fidelity}
\overline{F}=\int_{0}^{1}{F(\alpha^{2})d\alpha^{ 2}}=0.604.
\end{equation}

\section{ Broadcasting of quantum correlations using quantum copying machines}
In this section we study the feasibility of broadcasting of quantum correlations using the above mentioned quantum copying machines. We employ entanglement of formation and quantum discord for measuring the two aspects of the quantum correlations of  a two-qubit system, i.e., the quantum correlation defined  within the entanglement-separability paradigm and the one defined from an information-theoretic perspective. The general form of two-qubit density matrix $ \rho $ in Bloch representation is given by
\begin{equation}\label{state Bloch representation}
\rho=\frac{1}{4}\left(\mathbb{I}\otimes \mathbb{I}+\sum_{i=1}^{3}x_{i}\sigma _{i}\otimes \mathbb{I}+\sum_{i=1}^{3}y_{i} \mathbb{I}\otimes\sigma _{i}+\sum_{i,j=1}^{3}t_{i,j}\sigma _{i}\otimes\sigma_{j}\right),
\end{equation}
where $ x_{i}=\Tr\rho(\sigma _{i}\otimes \mathbb{I}) $, $ y_{i}=\Tr\rho(\mathbb{I}\otimes\sigma _{i}) $ are components of the local Bloch vectors, $ t_{i,j}=\Tr\rho(\sigma _{i}\otimes \sigma_{j}) $ are components of the correlation matrix $ T $.
By performing the following general quantum cloning machine
 \begin{eqnarray}\label{sym unitary transformation111}
U\ket{0}\ket{0}\ket{X}=\vert a\vert \ket{0}\ket{0}\ket{A}+\vert b\vert (\ket{0}\ket{1}+\ket{1}\ket{0})\ket{A_{\bot}},\\\nonumber
U\ket{1}\ket{0}\ket{X}=\vert a\vert \ket{1}\ket{1}\ket{A_{\bot}}+\vert b\vert (\ket{1}\ket{0}+\ket{0}\ket{1})\ket{A},
\end{eqnarray}
on the general two-qubit state introduced in Eq. \eqref{state Bloch representation}, we see that the Bloch vectors of output state are  $ \vec{x^{\prime}}=\lbrace 2\vert a\vert\vert b\vert x_{1},2\vert a\vert\vert b\vert x_{2},\vert a\vert^{2}x_{3}\rbrace $, $ \vec{y^{\prime}}=\lbrace 2\vert a\vert\vert b\vert y_{1},2\vert a\vert\vert b\vert y_{2},\vert a\vert^{2}y_{3}\rbrace $ and the components of the correlation matrix are $ \lbrace t^{\prime}_{11}=4\vert a\vert^{2}\vert b\vert^{2}t_{11}, t^{\prime}_{12}=4\vert a\vert^{2}\vert b\vert^{2}t_{12}, t^{\prime}_{13}=2\vert a\vert^{3}\vert b\vert t_{13}, t^{\prime}_{21}=4\vert a\vert^{2}\vert b\vert^{2}t_{21}, t^{\prime}_{22}=4\vert a\vert^{2}\vert b\vert^{2}t_{22}, t^{\prime}_{23}=2\vert a\vert^{3}\vert b\vert t_{23}, t^{\prime}_{31}=2\vert a\vert^{3}\vert b\vert t_{31}, t^{\prime}_{32}=2\vert a\vert^{3}\vert b\vert t_{32}, t^{\prime}_{33}=\vert a\vert^{4}t_{33}\rbrace $. 
Therefore, for our universal QCM defined by eq. \eqref{opt unitary transformation} we have $|a|^2=|b|=\frac{1}{2}$ and hence $\vec{x'}=(x'_1=\frac{1}{\sqrt{2}} x_1, x'_2=\frac{1}{\sqrt{2}} x_2, x'_3=\frac{1}{2} x_3)$, $\vec{y'}=(x'_1=\frac{1}{\sqrt{2}} y_1, y'_2=\frac{1}{\sqrt{2}} y_2, y'_3=\frac{1}{2} y_3)$ and $\lbrace t'_{11}=\frac{1}{2}t_{11},t'_{12}=\frac{1}{2}t_{12},t'_{13}=\frac{1}{2\sqrt{2}} t_{13},t'_{21}=\frac{1}{2}t_{21}, t'_{22}=\frac{1}{2}t_{22}, t'_{23}=\frac{1}{2\sqrt{2}} t_{23}, t'_{31}=\frac{1}{2\sqrt{2}} t_{31}, t'_{32}=\frac{1}{2\sqrt{2}} t_{32},t'_{33}=\frac{1}{4} t_{33}\rbrace$. Furthermore performing the non-local B-H QCM on the general two-qubit state \eqref{state Bloch representation} as input state, we obtain $ \lbrace \frac{6}{10}\vec{x}, \frac{6}{10}\vec{y}, \frac{6}{10}T\rbrace $.
For local B-H QCM, we have $ \vert a\vert^{2}=\frac{2}{3}, \vert b\vert^{2}=\frac{1}{6} $, so the Bloch vectors and the correlation matrix of output state are $ \lbrace \frac{2}{3}\vec{x}, \frac{2}{3}\vec{y}, \frac{4}{9}T\rbrace $. Also for two-Pauli-like cloning machine, we have $ \vert a\vert^{2}=\frac{4+\sqrt{79}}{21}, \vert b\vert^{2}=\frac{17-\sqrt{79}}{42} $, so the Bloch vectors of output state are  $ \vec{x^{\prime}}=\lbrace 0.6886 x_{1},0.6886 x_{2},0.6137x_{3}\rbrace $, $ \vec{y^{\prime}}=\lbrace 0.6886 y_{1},0.6886 y_{2},0.6137y_{3}\rbrace $ and the components of the correlation matrix are $ \lbrace t^{\prime}_{11}=0.4741t_{11}, t^{\prime}_{12}=0.4741t_{12}, t^{\prime}_{13}=0.4226 t_{13}, t^{\prime}_{21}=0.4741t_{21}, t^{\prime}_{22}=0.4741t_{22}, t^{\prime}_{23}=0.4226 t_{23}, t^{\prime}_{31}=0.4226 t_{31}, t^{\prime}_{32}=0.4226t_{32}, t^{\prime}_{33}=0.3766t_{33}\rbrace $.
The unitary transformation defined in eq. \eqref{sym unitary transformation111} maps $X$-states to $X$-states and hence all above mentioned QCMs preserve the $X$-shape of input state. So, for pure and Werner two-qubit input states, the output states of all above mentioned QCMs belong to the class of  $X$-states, i.e., their  density matrices take the following form in the standard basis
\begin{equation}
\rho_{AB} =\left(
\begin{array}{cccc}\rho_{11} & 0 & 0 & \rho_{14} \\
0 & \rho_{22} & \rho_{23} & 0 \\
0 & \rho_{32} & \rho_{33} & 0 \\
\rho_{41} & 0 & 0 & \rho_{44}
\end{array}\right).
\end{equation}
Here, for output states of all above QCMs we have $ \rho_{14}=\rho_{41} $ and $ \rho_{23}=\rho_{32} $.  
The entanglement of formation of a  general two-qubit state can be expressed as \cite{Wootters1998}
\begin{equation}\label{Eof}
E(\rho)=h\left( \frac{1}{2}+\frac{1}{2}\sqrt{1-C^{2}(\rho)}\right),
\end{equation}
where $ h(x)=-x\log_{2}x-(1-x)\log_{2}(1-x) $ is the Shannon entropy and  $ C(\rho) $ is the concurrence of the state.  For the class of $X$-states, considered above, $C(\rho)$ is given by 
\begin{equation}\label{concurrence}
C(\rho)=\max \left\lbrace 2\max\left\lbrace\mu_{1},\mu_{2},\mu_{3},\mu_{4}\right\rbrace -\mu_{1}-\mu_{2}-\mu_{3}-\mu_{4},0\right\rbrace,
\end{equation}
where
\begin{eqnarray}
\mu_{1,2}=\sqrt{\rho_{11}\rho_{44}}\pm |\rho_{14}|\\ \nonumber
\mu_{3,4}=\sqrt{\rho_{22}\rho_{33}}\pm |\rho_{23}|\\\nonumber
\end{eqnarray}
On the other hand, quantum discord of the $X$-state, when $ \rho_{14},\rho_{23}\in \mathbb{R}$, is given by \cite{Chen2011}
\begin{equation}
D_B(\rho)=S(\rho_{B})-S(\rho)+\min\{C_{1},C_{2}\},
\end{equation}
where $S(\rho)$ and $S(\rho_B)$ are von Neumann entropy of  $\rho$ and its reduced state $\rho_B$, respectively. Moreover, $C_1$ and $C_2$ are defined respectively by
\begin{align*}
C_{1} & =-\rho_{11}\log_{2}\left(\frac{\rho_{11}}{\rho_{11}+\rho_{33}}\right)-\rho_{22}\log_{2}\left(\frac{\rho_{22}}{\rho_{22}+\rho_{44}}\right)\\
 & \;\;\;-\rho_{33}\log_{2}\left(\frac{\rho_{33}}{\rho_{11}+\rho_{33}}\right)-\rho_{44}\log_{2}\left(\frac{\rho_{44}}{\rho_{22}+\rho_{44}}\right),
\end{align*}
and 
\[
C_{2}=-\left(\frac{1+\Upsilon}{2}\right)\log_{2}\left(\frac{1+\Upsilon}{2}\right)-\left(\frac{1-\Upsilon}{2}\right)\log_{2}\left(\frac{1-\Upsilon}{2}\right),
\]
where $\Upsilon=[(\rho_{11}+\rho_{22}-\rho_{33}-\rho_{44})^2+4(|\rho_{14}|+|\rho_{23}|)^{2}]^{\frac{1}{2}}$.

In the following we consider pure and Werner states as the two particular input states  of the above mentioned cloning machines.

\begin{figure}[ht!]
\centering
\includegraphics[width=15cm]{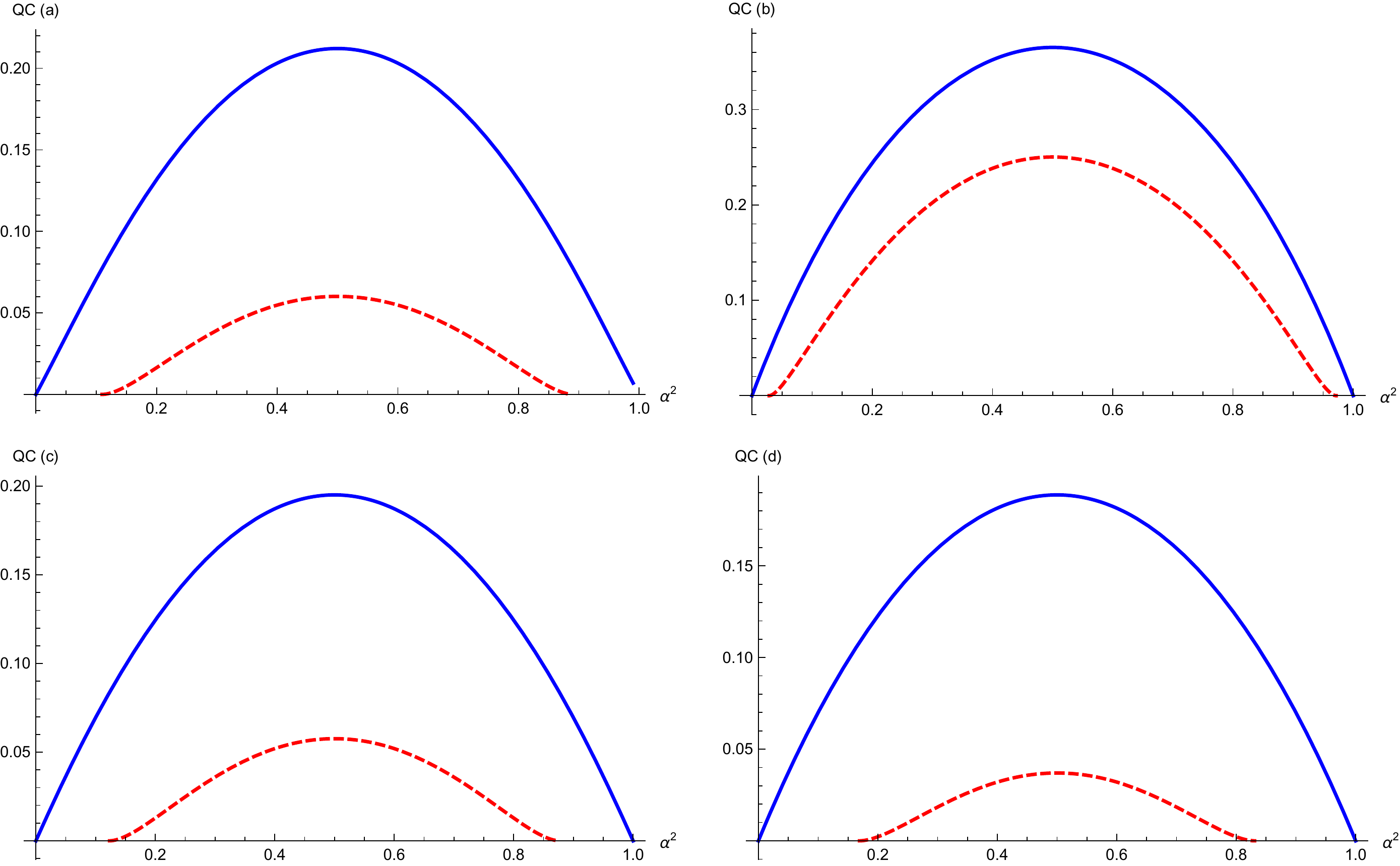}
\caption{(Color online) Quantum correlation of output states in terms of parameter $\alpha^{2}$ when the input state is pure for: a) local B-H  QCM , b)  non-local B-H QCM , c) two-Pauli-like QCM and d) UQCM defined in eq. \eqref{opt unitary transformation}. Blue solid lines show the quantum discord and the red dashed lines show the entanglement of formation.}
\label{figure1}
\end{figure}

{\it Two-qubit pure states.---}
For the two-qubit pure input state $ \ket{\psi}=\alpha\ket{00}+\beta\ket{11} $, the output states of the above cloning machines has $ X $-shape and the calculation of quantum discord and entanglement of formation  become simplified. The results depicted in Fig.  \ref{figure1} for various QCM outputs.
This figure shows that the non-local B-H QCM preserves more quantum discord and quantum entanglement during broadcasting procedure than other three machines, and the other QCMs behave almost similar for broadcasting the quantum discord. The results reveal that the quantum entanglement is more fragile than quantum discord during broadcasting procedure and it does not broadcast for some intervals of $ \alpha $.

\begin{figure}[ht!]
\centering
\includegraphics[width=15cm]{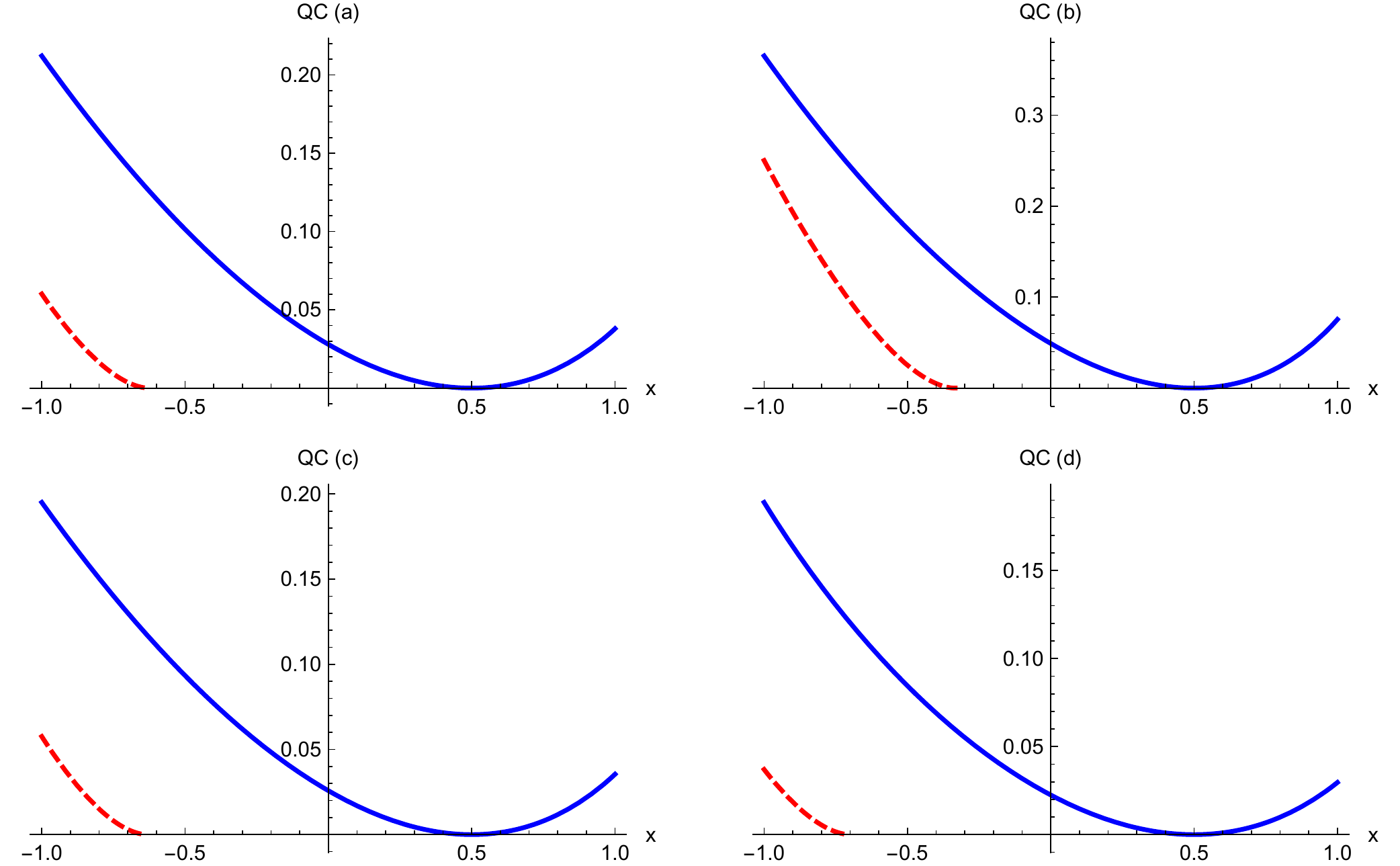}
\caption{(Color online) Quantum correlation in terms of the parameter $ x $  of the  Werner input state  for: a) local B-H QCM, b) non-local B-H QCM, c) two-Pauli-like QCM, and d) UQCM defined in eq. \eqref{opt unitary transformation}. Blue solid lines show quantum discord and the red dashed lines show EoF.}
\label{figure2}
\end{figure}

{\it Two-qubit Werner states.---}
Now, we consider the two-qubit Werner state, as input state. The Werner state is defined by
 \begin{equation}\label{Werner state}
\rho_{w} =\frac{(2-x)}{6}I+(\frac{2x-1}{6})\textbf{F},\quad x\in [-1,1],
\end{equation}
with $ \textbf{F}=\sum_{k,l=0}^{1}(\ket{kl}\bra{lk}) $. The Bloch vectors of Werner states are $ \vec{x}=0 $, $ \vec{y}=0 $ and the components of correlation matrix are $  t_{1}=t_{2}=t_{3}=\frac{1}{3}(2x-1) $. It is easy to show that the local B-H QCM changes the components of correlation matrix to $ t^{\prime}_{i}=\frac{4}{9}t_{i} $, the non-local B-H QCM changes them to  $ t^{\prime}_{i}=\frac{6}{10}t_{i} $, our UQCM introduced by Eq. \eqref{opt unitary transformation} changes them to $t^{\prime}_{1}=t^{\prime}_{2}=\frac{1}{2}t_{i}> t^{\prime}_{3}=\frac{1}{4}t_{i} $ and our two-Pauli-like QCM maps $t_i$s to  $ t^{\prime}_{1}=t^{\prime}_{2}=4\vert a\vert ^{2}\vert b\vert ^{2}t_{i}> t^{\prime}_{3}=\vert a\vert ^{4}t_{i} $. Obviously,  the first two QCMs transform  Werner states to Werner sates but the third and fourth one  transform  Werner state to a Bell-diagonal state. In this particular case quantum discord of the output states of first two QCMs can be written as \cite{Akhtarshenas2015}
 \begin{equation}\label{Werner discord}
D_{B}(\rho_{\out})=\frac{(1+t)}{4}\log(1+t)+\frac{(1-3t)}{4}\log(1-3t)-\frac{(1-t)}{2}\log(1-t),
\end{equation}
where $ t= t'_{1}=t'_{2}=t'_{3} $. Also in these cases concurrence of the output state is given by
\begin{equation}\label{Werner concurance}
C(\rho_{\out})=\max\left\lbrace 0,\frac{1}{4}\left(\sqrt{(3t-1)^{2}}-3\sqrt{(t+1)^{2}}\right)\right\rbrace,
\end{equation}
for $x\in [-1,0.5)$ and it is zero for $  x\in [0.5,1] $. Quantum discord of the output of our UQCM and two-Pauli-like QCM can be expressed as
\begin{eqnarray}\label{2pauli Werner discord}
D_{B}(\rho_{\out})&=&\frac{(1 + 2 t^{\prime}_{1} - t^{\prime}_{3})}{4}\log(1 + 2 t^{\prime}_{1} - t^{\prime}_{3})+\frac{(1 - 2 t^{\prime}_{1} - t^{\prime}_{3})}{4}\log(1 - 2 t^{\prime}_{1} - t^{\prime}_{3})\\\nonumber
&+&\frac{(1+t^{\prime}_{3})}{2}\log(1+t^{\prime}_{3})
-\frac{(1+t^{\prime}_{1})}{2}\log(1+t^{\prime}_{1})-\frac{(1-t^{\prime}_{1})}{2}\log(1-t^{\prime}_{1}),
\end{eqnarray}
and their concurrence is
\begin{equation}\label{2pauli Werner concurance}
C(\rho_{\out})=\max\left\lbrace 0,\frac{1}{4}\left(\sqrt{(2t^{\prime}_{1}+t^{\prime}_{3}-1)^{2}}-\sqrt{(2t^{\prime}_{1}+t^{\prime}_{3}+1)^{2}}
-2\sqrt{(t^{\prime}_{3}+1)^{2}}\right)\right\rbrace,
\end{equation}
for $x\in [-1,0.5)$.
Figure \ref{figure2} compares the results of all above mentioned QCMs.
This figure reveals that when the input state is Werner state, similar to pure input state, the non-local B-H QCM is more powerful for broadcasting the quantum discord and entanglement. Although quantum discord broadcasts properly for all values of $x$, quantum entanglement broadcasting is possible only for an small interval of $x$.

\section{Conclusion}
In this paper a class of universal quantum cloning machine has been presented for copying two-qubit pure states. Also with the aid of Pauli channel operations, two classes of state-dependent quantum copying machines were proposed for local cloning of two-qubit states. Furthermore, local broadcasting of quantum correlations, measured by quantum discord and entanglement of formation, via these machines have been studied for two-qubit pure and Werner states as input state. The results have been compared with the local and non-local Buzek-Hillery quantum cloning machines. The results show that quantum discord is more robust than entanglement of formation, during copying process. Moreover, non-local Buzek-Hillery machine preserves more quantum correlation than other machines. Since quantum information processing seeks locally controllable procedures, our local cloning machines provide a useful tools for operational broadcasting of quantum correlations.
\acknowledgments
The authors wish to thank The Office of Graduate
Studies of The University of Isfahan for their support.

\appendix
\section{Optimization for state-independent  cloning machine }\label{AppendixOptimizationuniversal}
In this appendix we provide procedure of maximization of the fidelity for state-independent  cloning machine. The maximization of the fidelity is done by the use of  Lagrange multiplier method.
We tend to
\begin{eqnarray}\label{constraints}
\textrm{maximize}\,\,F&=&(\vert a\vert^{2}+\vert b\vert^{2})^{2}\quad  \textrm{subject to} \\\nonumber
\phi_{1}&=&\vert a\vert^{2}+2\vert b\vert^{2}+\vert c\vert^{2}-1 \\\nonumber
\phi_{2}&=&\vert a\vert\vert c\vert\left(\langle A^{\prime}\vert C\rangle +\langle C\vert A^{\prime}\rangle \right)+2\vert b\vert^{2}\langle B\vert B^{\prime}\rangle \\\nonumber
\phi_{3}&=&2\left(\vert a\vert^{2}+\vert b\vert^{2}\right)-1-\vert a\vert^{2}\vert b\vert^{2}Re\left(\langle A\vert B^{\prime}\rangle +\langle  B\vert A^{\prime}\rangle \right)^{2}\\\nonumber
\phi_{4}&=&\vert a\vert^{2}\vert b\vert^{2}Im\left(\langle A\vert B^{\prime}\rangle +\langle  B\vert A^{\prime}\rangle \right)^{2} \\\nonumber
\phi_{5}&=&\vert b\vert^{2}\vert c\vert^{2}Re\left(\langle C\vert B^{\prime}\rangle +\langle B\vert C^{\prime}\rangle \right)^{2} \\\nonumber
\phi_{6}&=&\vert b\vert\left(\vert a\vert\langle B\vert A\rangle +\vert c\vert\langle C\vert B\rangle\right)\\\nonumber
\phi_{7}&=&\vert b\vert\left(\vert a\vert\langle A^{\prime}\vert B^{\prime}\rangle +\vert c\vert\langle B^{\prime}\vert C^{\prime}\rangle\right).
\end{eqnarray}
Using the Lagrange multiplier method we have:
\begin{eqnarray}\label{Lagrange multiplier}
\frac{\partial F}{\partial\vert a\vert}+\sum \lambda_{i}\frac{\partial \phi_{i}}{\partial\vert a\vert}=0\quad\\\nonumber
\frac{\partial F}{\partial\vert b\vert}+\sum \lambda_{i}\frac{\partial \phi_{i}}{\partial\vert b\vert}=0\quad\\\nonumber
\\\nonumber
\cdots\\\nonumber
\\\nonumber
\phi_{j}\equiv 0,\quad \quad j=1,\cdots 7,
\end{eqnarray}
where $ \lambda_{i} $ are the Lagrange multipliers.
From the partial derivative with respect to $ \vert c\vert $,
and taking the partial derivative with respect to $ \vert\langle A^{\prime}\vert C\rangle\vert $, $ \vert\langle C\vert A^{\prime}\rangle\vert $,  $ \vert\langle B^{\prime}\vert C\rangle\vert $, $ \vert\langle B\vert C^{\prime}\rangle\vert $, $ \vert\langle B^{\prime}\vert C^{\prime}\rangle\vert $, we find \begin{equation}\label{lambda1I}
2\vert c\vert^{2}\lambda_{1}=0.
\end{equation}
Also taking the partial derivative with respect to $ \vert\langle A^{\prime}\vert C\rangle\vert $, $ \vert\langle A^{\prime}\vert B\rangle\vert $, $ \vert\langle A\vert B\rangle\vert $, $ \vert\langle A^{\prime}\vert B^{\prime}\rangle\vert $ and $ \vert a\vert $, gives
\begin{equation}\label{lambda3}
4\vert a\vert\left(\vert a\vert ^{2}+\vert b\vert ^{2}-\lambda_{3}\right)+2\vert a\vert \lambda_{1}=0.
\end{equation}
From Eq. \eqref{lambda1I} we have
\begin{equation}\label{lambda3c}
\vert a\vert\vert c\vert \left(\vert a\vert ^{2}+\vert b\vert ^{2}-\lambda_{3}\right)=0.
\end{equation}
We want to obtain $ F\neq 0 $ hence the coefficients $ \vert a\vert $ and $ \vert b\vert $ can not be zero, simultaneously. In addition Eq. \eqref{lambda3c} implies that $ \lambda_{3}=\vert a\vert ^{2}+\vert b\vert ^{2} $ or $ \vert c\vert =0 $.\\
Now, we prove that if $ \lambda_{3}=\vert a\vert ^{2}+\vert b\vert ^{2} $ then $ F=0 $. We define $ \chi=\vert\chi\vert e^{ i\delta_{\chi}}=\langle B^{\prime}\vert A\rangle +\langle A^{\prime}\vert B\rangle $. Taking the partial derivative with respect to $ \vert\chi\vert $ and $ \delta_{\chi} $ leads to
\begin{equation}\label{chi}
-2\vert a\vert^{2}\vert b\vert^{2}\vert\chi\vert\cos 2\delta_{\chi}\lambda_{3}+2\vert a\vert^{2}\vert b\vert^{2}\vert\chi\vert\sin 2\delta_{\chi}\lambda_{4}=0,
\end{equation}
\begin{equation}\label{deltachi}
2\vert a\vert^{2}\vert b\vert^{2}\vert\chi\vert\sin 2\delta_{\chi}\lambda_{3}+2\vert a\vert^{2}\vert b\vert^{2}\vert\chi\vert\cos 2\delta_{\chi}\lambda_{4}=0.
\end{equation}
Inserting Eq.\eqref{chi} times $ \cos 2\delta_{\chi} $ to Eq.\eqref{deltachi} multiplied by $ \sin 2\delta_{\chi} $ gives
\begin{equation}\label{deltaApB}
\lambda_{3}\vert\chi\vert =0.
\end{equation}
So, if $ \lambda_{3}=\vert a\vert ^{2}+\vert b\vert ^{2} $ \textit{i.e.} $ \vert\chi\vert =0 $, then $ F=0 $. Therefore, we have $ \vert c\vert=0 $. Furthermore, from $ \phi_{2}\equiv 0 $, $ \phi_{6}\equiv 0 $ and $ \phi_{7}\equiv 0 $ we have
\begin{equation}\label{ABBpB}
\langle B\vert B^{\prime}\rangle =0, \quad \quad \langle A\vert B\rangle =0,
\quad \quad \langle A^{\prime}\vert B^{\prime}\rangle =0.
\end{equation}
Since $ \phi_{3}\equiv 0 $, we can write $ 2(\vert a\vert^{2}+\vert b\vert^{2})-1=\vert a\vert^{2}\vert b\vert^{2}\mu $, where $ \mu=Re\left(\langle A\vert B^{\prime}\rangle +\langle  B\vert A^{\prime}\rangle \right)^{2} $. Fidelity $ F $ is maximum if $ \mu $ takes its maximum value, that is $ \mu=4 $, therefore
\begin{equation}\label{valuab}
\vert a\vert=\frac{1}{\sqrt{2}}, \quad \quad \vert b\vert=\frac{1}{2}.
\end{equation}
In summary, we arrive at  a class of optimal universal quantum cloning machine for local cloning of two-qubit pure states as
 \begin{eqnarray}\label{opt unitary transformation1}
U\ket{0}\ket{0}\ket{X}=\frac{1}{\sqrt{2}} \ket{0}\ket{0}\ket{A}+\frac{1}{2} \left(\ket{0}\ket{1}+\ket{1}\ket{0}\right)\ket{A_{\perp}}\\\nonumber
U\ket{1}\ket{0}\ket{X}=\frac{1}{\sqrt{2}} \ket{1}\ket{1}\ket{A_{\perp}}+\frac{1}{2} \left(\ket{1}\ket{0}+\ket{0}\ket{1}\right)\ket{A}.
\end{eqnarray}

\section{Optimization for state-dependent cloning machine }
\label{AppendixOptimizationdependents tate}
In the case of state-dependent cloning machine we require to
 \begin{eqnarray}\label{constraints2}
\textrm{maximize}\,\, s&=&\frac{1}{5}(8(\vert a\vert^{2}+\vert b\vert^{2})^{2}-3)\quad \textrm{subject to} \\\nonumber
\phi_{1}&=&\vert a\vert^{2}+2\vert b\vert^{2}+\vert c\vert^{2}-1\\\nonumber
\phi_{2}&=&\vert a\vert\vert c\vert(\langle A^{\prime}\vert C\rangle +\langle A\vert C^{\prime}\rangle )+2\vert b\vert^{2}\langle B\vert B^{\prime}\rangle \\\nonumber
\phi_{3}&=&\frac{1}{5}(8(\vert a\vert^{2}+\vert b\vert^{2})^{2}-28)+8(\vert a\vert^{2}+\vert b\vert^{2})-4\vert a\vert^{2}\vert b\vert^{2}Re(\langle A\vert B^{\prime}\rangle +\langle  B\vert A^{\prime}\rangle )^{2}\\\nonumber
\phi_{4}&=&4\vert a\vert^{2}\vert b\vert^{2}Im(\langle A\vert B^{\prime}\rangle +\langle  B\vert A^{\prime}\rangle )^{2} \\\nonumber
\phi_{5}&=&\vert b\vert^{2}\vert c\vert^{2}Re(\langle C\vert B^{\prime}\rangle +\langle B\vert C^{\prime}\rangle )^{2} \\\nonumber
\phi_{6}&=&\vert b\vert(\vert a\vert\langle B\vert A\rangle +\vert c\vert\langle C\vert B\rangle) \\\nonumber
\phi_{7}&=&\vert b\vert(\vert a\vert\langle A^{\prime}\vert B^{\prime}\rangle +\vert c\vert\langle B^{\prime}\vert C^{\prime}\rangle)
\end{eqnarray}
Using the Lagrange multiplier method and following the procedure of previous appendix leads to
\begin{equation}
\vert a\vert ^{2}=\frac{4+\sqrt{79}}{21},\quad\quad \vert b\vert ^{2}=\frac{17-\sqrt{79}}{42}, \quad\quad \vert c\vert=0.
\end{equation}
In Summary, we have a class of state-dependent quantum cloning machine ( two-Pauli-like cloning machine) for local cloning of two-qubit states as
 \begin{eqnarray}\label{two-Pauli-like transformation}
U\ket{0}\ket{0}\ket{X}=\sqrt{\frac{4+\sqrt{79}}{21}} \ket{0}\ket{0}\ket{A}+\sqrt{\frac{17-\sqrt{79}}{42}} (\ket{0}\ket{1}+\ket{1}\ket{0})\ket{A_{\perp}},\\\nonumber
U\ket{1}\ket{0}\ket{X}=\sqrt{\frac{4+\sqrt{79}}{21}} \ket{1}\ket{1}\ket{A_{\perp}}+\sqrt{\frac{17-\sqrt{79}}{42}} (\ket{1}\ket{0}+\ket{0}\ket{1})\ket{A}.
\end{eqnarray}

\end{document}